\documentclass{article}

\usepackage[final]{neurips_2025}

\usepackage[utf8]{inputenc} 
\usepackage[T1]{fontenc}    

\usepackage{hyperref}       
\usepackage{url}            
\usepackage{microtype}      
\usepackage{parskip}        
\usepackage{nicefrac}       
\usepackage{graphicx}       

\usepackage{amsmath}        
\usepackage{amssymb}        
\usepackage{amsfonts}       

\usepackage[table]{xcolor}  
\usepackage{booktabs}       
\usepackage{tabularx}       
\usepackage{siunitx}        

\usepackage{tcolorbox}      
\tcbuselibrary{skins}       
\usepackage{float}

\title{An Evaluation of Interleaved Instruction Tuning on Semantic Reasoning Performance in an Audio MLLM}

\author{%
  Jiawei Liu\textsuperscript{1}
  \quad
  \textbf{Enis Berk Çoban}\textsuperscript{1}
  \quad
  \textbf{Zarina Schevchenko}\textsuperscript{2}
  \quad
  \textbf{Hao Tang}\textsuperscript{1,3}
  \quad
  \textbf{Zhigang Zhu}\textsuperscript{1,4}
  \and
  \textbf{Michael I Mandel}\textsuperscript{1}
  \quad
  \textbf{Johanna Devaney}\textsuperscript{1,2}
  \\[0.25em]
  \centering
  \textsuperscript{1}The Graduate Center, CUNY \quad
  \textsuperscript{2} Brooklyn College, CUNY 
  \quad
  \and
  \textsuperscript{3} 
  Borough of Manhattan Community College, CUNY \quad
  \textsuperscript{4} 
  The City College of New York, CUNY
  \quad
}

\begin{document}

\maketitle

\begin{abstract}
  Standard training for Multi-modal Large Language Models (MLLMs) involves concatenating non-textual information, like vision or audio, with a text prompt. This approach may not encourage deep integration of modalities, limiting the model's ability to leverage the core language model's reasoning capabilities. This work examined the impact of interleaved instruction tuning in an audio MLLM, where audio tokens are interleaved within the prompt. Using the Listen, Think, and Understand (LTU) model as a testbed, we conduct an experiment using the Synonym and Hypernym Audio Reasoning Dataset (SHARD), our newly created reasoning benchmark for audio-based semantic reasoning focusing on synonym and hypernym recognition. Our findings show that while even zero-shot interleaved prompting improves performance on our reasoning tasks, a small amount of fine-tuning using interleaved training prompts improves the results further, however, at the expense of the MLLM's audio labeling ability.
\end{abstract}

\section{Introduction}
Multi-modal Large Language Models (MLLMs) continue to face significant reasoning challenges. In the visual domain, models often struggle with complex spatial relationships and compositional understanding \cite{chen2024spatialvlm, cheng2024spatialrgpt, ma2024spatialpin, tong2024eyes}. In the audio domain, the challenges faced with visual signals persist and are compounded by the inherently temporal nature of audio signals. 
Our prior work \cite{ccoban2024mllms} demonstrated that audio MLLMs fail to connect the LLM's text-based reasoning abilities to audio modality, particularly struggling with semantic reasoning tasks.
Recent studies on Vision-Language Models (VLMs) have suggested that some models behave like "bag-of-words" encoders, focusing on a few salient features rather than processing the input sequentially and contextually \cite{yuksekgonul2022and, qi2025beyond}. This suggests that model may not be "reading" and reasoning over the non-textual input with the same attention it applies to text. To address this limitation, researchers have begun used interleaved training prompts, where multiple modalities like text, images, and audio are interwoven in a single, continuous sequence \cite{wang2024emu3, zhan2024anygpt, huang2023language, li2024llava, li2025otter, feofanov2025mantis}. The hypothesis is that by positioning non-text tokens amongst text tokens within the sequence, the LLM's core attention mechanisms can be leveraged to encourage it to pay closer attention to the multi-modal inputs and their relationship to text. 

To address the semantic reasoning gap identified in our previous study \cite{ccoban2024mllms}, we present a systematic investigation of the impact of interleaved prompts for similarity-based and hierarchical semantic reasoning with audio. We introduce a new benchmark designed to evaluate audio-based semantic reasoning through two tasks: synonym (similarity) and hypernym (hierarchical) reasoning. Our study follows a three-stage approach. Firstly, we examine the zero-shot impact of applying the interleaved prompts to the original Listen, Think and Understand (LTU) model \cite{gong2023listen}. Second, we examine the impact of fine-tuning the model with an interleaved instruction dataset with both small-scale (40K samples) and large-scale (1 million samples) training sets. We also evaluate the impact of our fine-tuning on the MLLM's audio labeling functionality.

\section{Background}


MLLMs are built to integrate non-textual modalities, like vision and audio, with the powerful reasoning abilities of LLMs. While visual MLLMs show promise, research has revealed significant reasoning limitations, such as poor spatial reasoning \cite{qi2025beyond, chen2024spatialvlm, ma2024spatialpin} and a tendency for catastrophic forgetting of the core LLM's capabilities when non-linguistic prompts are introduced \cite{wang2023paxion}.
This reasoning gap also persists in recent audio MLLMs. Models such as LTU \cite{gong2023listen} and SALMONN \cite{tang2023salmonn} have been developed to answer questions about audio. However, our prior work \cite{ccoban2024mllms} demonstrated that while these two models are proficient at descriptive tasks like audio captioning, they cannot fully leverage their LLM's text-based reasoning for complex semantic tasks (specifically, synonym and hypernym identification). This finding suggests that the standard training method creates a shallow integration between the LLM and the multi-modal encoder, limiting the MLLM's ability to apply the LLM's conceptual knowledge directly to the audio input.

One way to address this integration failure is interleaved instruction tuning, where audio tokens are interleaved directly within the text prompt. 
This method has been applied in training various vision and audio MLLMs novel architectures (e.g., Chameleon \cite{team2024chameleon}, Mantis \cite{feofanov2025mantis}, Emu3 \cite{wang2024emu3}, and APT \cite{liang2025acoustic}), however, none of these papers have systematically examined the specific impact of interleaving itself. In this work, we investigate the impact of interleaving audio tokens within the text prompt on the LTU \cite{gong2023listen} model's semantic reasoning capabilities.

\section{Method}

\paragraph{Base Model and Interleaved Adaptation.}

Our investigation builds on the LTU model, which pairs an audio encoder $E_A$ using CAV-MAE \cite{gong2022contrastive} and a LLaMa \cite{touvron2023llama} backbone. The audio encoder $E_A$ processes 10-second raw audio clip $A_{clip}$ into 32 embeddings $H_A \in \mathbb{R}^{32 \times d_{audio}}$ and projects them to a 4096-dimensional space $Z_A \in \mathbb{R}^{32 \times 4096}$ to match the LLM's input. The text prompt $T$ is tokenized and embedded by the LLM's embedding layer to obtain $E_T \in \mathbb{R}^{L \times 4096}$, where $L$ is the number of text tokens. For the conventional non-interleaved setup, also used in LTU, it typically prepends the entire audio embeddings $Z_A$ to the text embeddings:
$$
I_{\text{non-interleaved}} = [Z_A ; E_T] = \{z_1, \ldots, z_{32}, t_1, \ldots, t_L\}
$$
For an interleaved setup, we place the audio embeddings at any position of the text (e.g., between or end). If we place audio embeddings between of text, the prompt $T$ will be structured as $T = (T_{\text{pre}}, \langle\text{audio}\rangle, T_{\text{post}})$. Here, $\langle\text{audio}\rangle$ represents a conceptual placeholder for audio modality, $T_{pre}$ is the text before the audio, and $T_{post}$ is the text after. The final input sequence $I_{\text{interleaved}}$ is:
$$I_{\text{interleaved}} = [E_{\text{pre}} ; Z_A ; E_{\text{post}}] = \{t_1, \ldots, t_i, z_1, \ldots, z_{32}, t_{i+1}, \ldots, t_L\}$$

To adapt the original LTU for this interleaved structure, we implement this concept by introducing a new special token \texttt{[AUDIO]}. During the preprocessing, each occurrence of \texttt{[AUDIO]} expands to 32 positions. At runtime, those 32 positions are replaced by the audio embeddings $Z_A$, and the rest of the sequence consists of ordinary text tokens. The detailed mathematical formulation is listed in Appendix A. For parameter-efficient fine-tuning, we replicate LTU's Low-Rank Adaptation (LoRA) \cite{hu2022lora} setup with a rank of $r=8$ and scale $\alpha=16$. The training objective is standard autoregressive cross-entropy over the sequence. During the finetuning, we keep input data as training data but we mask the labels of audio tokens hence only text tokens contribute to the loss. All fine-tuning experiments are conducted on two A40 GPUs using a batch size of 4 and a learning rate of $2e-5$.

\paragraph{Fine-tuning Dataset.}
To study interleaved instruction tuning, we created a 1-million-item finetuning dataset (as well as a 40k-item subset) with interleaved prompt by reformulating the audio question-answering dataset released by LTU \cite{gong2023listen}, rather than introducing new audio sources or reasoning-related question-answering data. 
This approach ensures that any observed performance changes can be more directly attributed to the change in interleaved prompt adaptation. The released audio question-answering dataset contains a mix of closed-ended and open-ended questions and audio sources from various public benchmarks \cite{gemmeke2017audio, hershey2021benefit_audioset_strong, piczak2015esc, fonseca2021fsd50k}. In our experiment,  we used samples sourced from the AudioSet-Strong \cite{hershey2021benefit_audioset_strong} training set for open-ended questions and samples from both the AudioSet-Balanced  \cite{gemmeke2017audio} training and AudioSet-Strong \cite{hershey2021benefit_audioset_strong} training sets for closed-ended classification tasks. 
A core principle of our interleaved prompt design is to minimize explicit modality cues to encourage the LLM to process the audio embeddings more akin to textual tokens within the sequence. To achieve this, we omitted phrases that explicitly refer to the audio modality, such as "in this audio clip" or ``listen to this recording'' from the original non-interleaved prompt. To generate a diverse and linguistically rich dataset, we designed a series of new system and user prompt templates with GPT-4.1 mini to rephrase the original prompt into an interleaved format, embedding a special \texttt{[AUDIO]} placeholder within the prompt text. A temperature sampling strategy with values ranging from 0.7 to 1.1 was employed during this generation process to encourage variability in wording. Crucially, while the textual prompt was reformulated, the source audio clip and the corresponding ground-truth answer label remained identical to the original dataset, ensuring that the primary experimental variable was the prompt format itself. The complete prompt generation methodology, including all prompt templates and representative data samples, is detailed in Appendix B.
\paragraph{Audio Reasoning Benchmark.}
To evaluate the model's ability to apply lexcial reasoning to audio inputs, we developed the \underline{\textbf{S}}ynonym and \underline{\textbf{H}}ypernym \underline{\textbf{A}}udio \underline{\textbf{R}}easoning \underline{\textbf{D}}ataset
(SHARD)
building on our earlier work \cite{ccoban2024mllms}, focused on two fundamental semantic relationships, \textbf{synonym} and \textbf{hypernym}. It facilitates evaluation of whether an MLLM understands and uses the audio to support a textual relation decision, forming a solid audio reasoning benchmark.
Specifically, synonyms test the model's ability to recognize semantic equivalence (e.g., car and automobile), while hypernyms assess its grasp of hierarchical classification (e.g., recognizing that a car is a type of vehicle). 
SHARD consists of 78 words, with two synonyms and hypernyms per word and four audio examples of the word from the AudioSet evaluation dataset \cite{gemmeke2017audio}, selected to include only audio files that contain only the sound of the corresponding word/label. These labels were selected to ensure broad categorical coverage across three types of sounds found in AudioSet: anthrophony (sounds made by humans), biohphony (sounds made by animals), and geophony (sounds made by non-biological nature). 
Synonyms and hypernyms were collected from the lexical database WordNet \cite{fellbaum2010wordnet}.

\paragraph{Experimental Methodology.}
Our experiment consists of three stages. First, we investigate the zero-shot impact of applying interleaved prompts directly to the baseline LTU model. Second, we fine-tune the model on a subset (40K samples) of the full fine-tuning dataset described above. And third, we use the full fine-tuning dataset (1 million samples) to observe the impact of data volume on the task. 
For each stage, we evaluate the performance of model variants with the SHARD benchmark's prompts on synonym and hypernym based reasoning tasks. 
The specific prompt templates used to examine these relationships in both non-interleaved and interleaved formats are detailed in Table \ref{exp2prompts}.


Each unique query from SHARD (a specific audio file and word pair) was repeated four times to account for the stochastic nature of the model's responses.
To efficiently evaluate the model's responses to the synonym and hypernym tasks, we parse the generated text using whole-word regular expressions to extract a binary decision. We map common negative phrases (e.g., "no", "does not", "is not") to a 'No' decision and affirmative phrases (e.g., "yes", "does") to a 'Yes' decision.  We evaluate the models' ability to label audio events via regular expression exact match against the canonical AudioSet label strings.


\begin{table*}

\caption{Prompt templates for the Identity, Synonym, and Hypernym evaluation tasks.
For all \emph{Non-interleaved} prompts, the audio embeddings are automatically prepended to the textual instruction. The \texttt{[AUDIO]} token indicates where audio embeddings are inserted for the \emph{Interleaved} format.}

\centering

\begin{tabular}{llp{0.6\textwidth}} 


\toprule
 Identity & \emph{Non-interleaved} & Can you list the labels based on this audio file?\\
 & \emph{Interleaved} & Can you list the labels based on \texttt{[AUDIO]}?
 \\
 \midrule
 
 Similarity & \emph{Non-interleaved} & Is the sound of the object in this audio signal similar to \{synonym\}?\\

 (synonym) & \emph{Interleaved} & Is \texttt{[AUDIO]} similar to \{synonym\}?\\
 \midrule

 Hierarchy & \emph{Non-interleaved} & Is the sound of the object in this audio signal a type of \{hypernym\}?\\
(hypernym) & \emph{Interleaved} & Is \texttt{[AUDIO]} a type of \{hypernym\}?\\

 \bottomrule

\end{tabular}


\label{exp2prompts}
\vspace{-0.5cm}

\end{table*}

\section{Results}
%

We evaluated each model configuration on SHARD, reporting Accuracy, Precision, Recall, and F1 scores for both the synonym and hypernym tasks in Table \ref{tab:result}, along with the model's ability to predict the original AudioSet label (Identity).
Zero-shot interleaving on the original LTU model improved recall performance from 18.31\% to 54.13\% for synonyms and 52.76\% to 70.82\% for hypernyms. But this came at the expense of the precision, which declined from 96.45\% to 52.90\% for synonyms and 98.90\% to 61.33\% for hypernyms, compared to the non-interleaved prompting of the original LTU model. Overall, interleaved prompting increased the F1 score on both the synonym task from 38.78\% to 53.51\%, but it reduced the F1 score slightly on the hypernym task (from 68.81\% to 64.75\%) as well as accuracy on the identity task from 48.45\% to 41.69 \%. 

Fine-tuning with the small dataset, improved precision for both synonyms (84.81\%) and hypernyms (91.45\%) over the interleaved prompting of the baseline model, with only a small decline in recall, 44.40\% for synonyms and 66.18\% for hypernyms. This resulted in a higher F1 score for both the synonym (58.29\%) and hypernym (76.69\%) tasks over either prompting strategies on the baseline model. Accuracy on identity, however, declined to 39.33\%, lower than both prompting strategies. 

Fine-tuning with the large dataset (FT-Large) resulted in a sharp decline in recall compared to the FT-Small model, dropping to 6.31\% for synonyms and 38.67\% for hypernyms. At the same time, this model's precision was 100.00\% for synonyms and 87.10\% for hypernyms while identity accuracy fell to 28.75\%.

\section{Discussion}


Our results demonstrate that: (1) First, MLLMs with non-interleaved setup can benefit from interleaving. Even without training, simply changing the prompt from a non-interleaved to an interleaved format improved the baseline LTU model's behavior on semantic reasoning tasks, although the model's ability to label audio events deteriorated.
(2) Second, small-scale fine-tuning on interleaved prompts (FT-Small) achieves the best-balanced F1 scores for both synonyms and hypernyms. We hypothesize this improvement stems from the attention mechanism. The interleaved instruction allows audio tokens to attend directly to preceding text, rather than only to other audio tokens. This provides immediate textual context for the audio, forcing the model to establish a tighter, more granular alignment, which in turn improves its reasoning accuracy.
(3) Third, naively scaling the fine-tuning data (FT-Large) resulted in a significant performance collapse. We observe overfitting, where the model becomes overly conservative with decreasing recall and increasing precision on both the synonym and hypernym tasks. We also observed catastrophic forgetting through the decline in accuracy on the identity (labeling) tasks. The decline in labeling accuracy is likely contributing to the substantial decrease in recall for the synonym task, as there is an increase in misclassification with objects of similar type (e.g., sailboat and motorboat), which causes issues for the identity and synonym tasks but not the hypernyms one because they still fall into the same larger-scale category. 

\begin{table}[t]
\centering
\small
\setlength{\tabcolsep}{4pt}
\renewcommand{\arraystretch}{1.1}
\begin{tabular}{l *{9}{S[table-format=2.2]}}
\toprule
& \multicolumn{3}{c}{Accuacy (\%)} & \multicolumn{2}{c}{Precision (\%)} & \multicolumn{2}{c}{Recall (\%)} & \multicolumn{2}{c}{F1 Score (\%)} \\
\cmidrule(lr){2-4}\cmidrule(lr){5-6}\cmidrule(lr){7-8}\cmidrule(lr){9-10}
Model &
\multicolumn{1}{c}{Identity} & \multicolumn{1}{c}{Syno.} & \multicolumn{1}{c}{Hyper.} &
\multicolumn{1}{c}{Syno.} & \multicolumn{1}{c}{Hyper.} &
\multicolumn{1}{c}{Syno.} & \multicolumn{1}{c}{Hyper.} &
\multicolumn{1}{c}{Syno.} & \multicolumn{1}{c}{Hyper.} \\
\midrule
LTU baseline (non-interleaved) & \textbf{48.45} & 59.50 & 76.40 & 96.45 & \textbf{98.90} & 18.31 & 52.76 & 30.78 & 68.81 \\
LTU baseline(interleaved) & 41.69 & 53.75 & 63.57 & 52.90 & 61.33 & \textbf{54.13} & \textbf{70.82} & 53.51 & 65.74 \\
\hline
LTU FT--Small (interleaved) & 39.33 & \textbf{68.75} & \textbf{80.26} & 84.81 & 91.45 & 44.40 & 66.18 & \textbf{58.29} & \textbf{76.79} \\
\hline
LTU FT--Large (interleaved) & 28.75 & 53.93 & 66.91 & \textbf{100.00} & 87.10 & 6.31 & 38.67 & 11.88 & 53.56 \\
\bottomrule
\end{tabular}
\caption{Results on the identity, synonym (Syno.), and hypernym (Hyper.) tasks. Values are percentages for Accuracy, Precision, Recall, and F1. \textbf{LTU baseline} refers to vanilla LTU. \textbf{FT} denotes fine-tuned models; “Small/Large” indicate the size of the fine-tuning dataset. }
\vspace{-0.85cm}
\label{tab:result}
\end{table}

\section{Conclusions and Future Work}

In this work\footnote{The code is available at \href{https://github.com/LUMaA-CUNY/Interleaved-Audio-MLLM}{https://github.com/LUMaA-CUNY/Interleaved-Audio-MLLM} and SHARD is available at \href{https://github.com/LUMaA-CUNY/SHARD}{https://github.com/LUMaA-CUNY/SHARD}.}, we demonstrated the positive impact of interleaved instruction tuning on semantic reasoning capabilities in an audio-based MLLM.  We also introduced SHARD, a novel benchmark for audio-based semantic reasoning with synonyms and hypernyms.)
Through a series of experiments with SHARD, we showed that using interleaved prompting on the original MLLM improved semantic reasoning and that small-scale fine-tuning (FT-Small) on interleaved prompts improved results further. However, 
naively scaling the fine-tuning (FT-Large) can lead to overfitting and catastrophic forgetting of the MLLM's audio labeling capability. The decline in audio labeling capability, was observed to a lesser extent in all interleaved conditions,  suggesting that, in future work, incorporating diverse interleaved data earlier in the pre-training phase would facilitate more robust and generalizable audio-reasoning MLLMs.

\begin{ack}
We would like to acknowledge the contribution of 
Claire Amont Alexandre to data collection for SHARD.   

This material is based on work supported by the National Science Foundation under Grants Nos. 1839185, 1839198, and 2228910.
\end{ack}

\bibliographystyle{plain}
\bibliography{references}

\begin{thebibliography}{10}

\bibitem{chen2024spatialvlm}
Boyuan Chen, Zhuo Xu, Sean Kirmani, Brain Ichter, Dorsa Sadigh, Leonidas Guibas, and Fei Xia.
\newblock Spatialvlm: Endowing vision-language models with spatial reasoning capabilities.
\newblock In {\em Proceedings of the IEEE/CVF Conference on Computer Vision and Pattern Recognition}, pages 14455--14465, 2024.

\bibitem{cheng2024spatialrgpt}
An-Chieh Cheng, Hongxu Yin, Yang Fu, Qiushan Guo, Ruihan Yang, Jan Kautz, Xiaolong Wang, and Sifei Liu.
\newblock Spatialrgpt: Grounded spatial reasoning in vision-language models.
\newblock {\em Advances in Neural Information Processing Systems}, 37:135062--135093, 2024.

\bibitem{ccoban2024mllms}
EB~{\c{C}}oban, MI~Mandel, and J~Devaney.
\newblock What do mllms hear? examining the interaction between llm and audio encoder components in multimodal large language models.
\newblock In {\em Audio Imagination: NeurIPS 2024 Workshop AI-Driven Speech, Music, and Sound Generation}, 2024.

\bibitem{fellbaum2010wordnet}
Christiane Fellbaum.
\newblock Wordnet.
\newblock In {\em Theory and applications of ontology: computer applications}, pages 231--243. Springer, 2010.

\bibitem{feofanov2025mantis}
Vasilii Feofanov, Songkang Wen, Marius Alonso, Romain Ilbert, Hongbo Guo, Malik Tiomoko, Lujia Pan, Jianfeng Zhang, and Ievgen Redko.
\newblock Mantis: Lightweight calibrated foundation model for user-friendly time series classification.
\newblock {\em arXiv preprint arXiv:2502.15637}, 2025.

\bibitem{fonseca2021fsd50k}
Eduardo Fonseca, Xavier Favory, Jordi Pons, Frederic Font, and Xavier Serra.
\newblock Fsd50k: an open dataset of human-labeled sound events.
\newblock {\em IEEE/ACM Transactions on Audio, Speech, and Language Processing}, 30:829--852, 2021.

\bibitem{gemmeke2017audio}
Jort~F Gemmeke, Daniel~PW Ellis, Dylan Freedman, Aren Jansen, Wade Lawrence, R~Channing Moore, Manoj Plakal, and Marvin Ritter.
\newblock Audio set: An ontology and human-labeled dataset for audio events.
\newblock In {\em 2017 IEEE international conference on acoustics, speech and signal processing (ICASSP)}, pages 776--780. IEEE, 2017.

\bibitem{gong2023listen}
Yuan Gong, Hongyin Luo, Alexander~H Liu, Leonid Karlinsky, and James Glass.
\newblock Listen, think, and understand.
\newblock {\em arXiv preprint arXiv:2305.10790}, 2023.

\bibitem{gong2022contrastive}
Yuan Gong, Andrew Rouditchenko, Alexander~H Liu, David Harwath, Leonid Karlinsky, Hilde Kuehne, and James Glass.
\newblock Contrastive audio-visual masked autoencoder.
\newblock {\em arXiv preprint arXiv:2210.07839}, 2022.

\bibitem{hershey2021benefit_audioset_strong}
Shawn Hershey, Daniel~PW Ellis, Eduardo Fonseca, Aren Jansen, Caroline Liu, R~Channing Moore, and Manoj Plakal.
\newblock The benefit of temporally-strong labels in audio event classification.
\newblock In {\em ICASSP 2021-2021 IEEE International Conference on Acoustics, Speech and Signal Processing (ICASSP)}, pages 366--370. IEEE, 2021.

\bibitem{hu2022lora}
Edward~J Hu, Yelong Shen, Phillip Wallis, Zeyuan Allen-Zhu, Yuanzhi Li, Shean Wang, Lu~Wang, Weizhu Chen, et~al.
\newblock Lora: Low-rank adaptation of large language models.
\newblock {\em ICLR}, 1(2):3, 2022.

\bibitem{huang2023language}
Shaohan Huang, Li~Dong, Wenhui Wang, Yaru Hao, Saksham Singhal, Shuming Ma, Tengchao Lv, Lei Cui, Owais~Khan Mohammed, Barun Patra, et~al.
\newblock Language is not all you need: Aligning perception with language models.
\newblock {\em Advances in Neural Information Processing Systems}, 36:72096--72109, 2023.

\bibitem{li2025otter}
Bo~Li, Yuanhan Zhang, Liangyu Chen, Jinghao Wang, Fanyi Pu, Joshua~Adrian Cahyono, Jingkang Yang, Chunyuan Li, and Ziwei Liu.
\newblock Otter: A multi-modal model with in-context instruction tuning.
\newblock {\em IEEE Transactions on Pattern Analysis and Machine Intelligence}, 2025.

\bibitem{li2024llava}
Feng Li, Renrui Zhang, Hao Zhang, Yuanhan Zhang, Bo~Li, Wei Li, Zejun Ma, and Chunyuan Li.
\newblock Llava-next-interleave: Tackling multi-image, video, and 3d in large multimodal models.
\newblock {\em arXiv preprint arXiv:2407.07895}, 2024.

\bibitem{liang2025acoustic}
Jinhua Liang, Xubo Liu, Wenwu Wang, Mark~D Plumbley, Huy Phan, and Emmanouil Benetos.
\newblock Acoustic prompt tuning: Empowering large language models with audition capabilities.
\newblock {\em IEEE Transactions on Audio, Speech and Language Processing}, 2025.

\bibitem{ma2024spatialpin}
Chenyang Ma, Kai Lu, Ta-Ying Cheng, Niki Trigoni, and Andrew Markham.
\newblock Spatialpin: Enhancing spatial reasoning capabilities of vision-language models through prompting and interacting 3d priors.
\newblock {\em Advances in neural information processing systems}, 37:68803--68832, 2024.

\bibitem{piczak2015esc}
Karol~J Piczak.
\newblock Esc: Dataset for environmental sound classification.
\newblock In {\em Proceedings of the 23rd ACM international conference on Multimedia}, pages 1015--1018, 2015.

\bibitem{qi2025beyond}
Jianing Qi, Jiawei Liu, Hao Tang, and Zhigang Zhu.
\newblock Beyond semantics: Rediscovering spatial awareness in vision-language models.
\newblock {\em arXiv preprint arXiv:2503.17349}, 2025.

\bibitem{tang2023salmonn}
Changli Tang, Wenyi Yu, Guangzhi Sun, Xianzhao Chen, Tian Tan, Wei Li, Lu~Lu, Zejun Ma, and Chao Zhang.
\newblock Salmonn: Towards generic hearing abilities for large language models.
\newblock {\em arXiv preprint arXiv:2310.13289}, 2023.

\bibitem{team2024chameleon}
Chameleon Team.
\newblock Chameleon: Mixed-modal early-fusion foundation models.
\newblock {\em arXiv preprint arXiv:2405.09818}, 2024.

\bibitem{tong2024eyes}
Shengbang Tong, Zhuang Liu, Yuexiang Zhai, Yi~Ma, Yann LeCun, and Saining Xie.
\newblock Eyes wide shut? exploring the visual shortcomings of multimodal llms.
\newblock In {\em Proceedings of the IEEE/CVF Conference on Computer Vision and Pattern Recognition}, pages 9568--9578, 2024.

\bibitem{touvron2023llama}
Hugo Touvron, Thibaut Lavril, Gautier Izacard, Xavier Martinet, Marie-Anne Lachaux, Timoth{\'e}e Lacroix, Baptiste Rozi{\`e}re, Naman Goyal, Eric Hambro, Faisal Azhar, et~al.
\newblock Llama: Open and efficient foundation language models.
\newblock {\em arXiv preprint arXiv:2302.13971}, 2023.

\bibitem{wang2024emu3}
Xinlong Wang, Xiaosong Zhang, Zhengxiong Luo, Quan Sun, Yufeng Cui, Jinsheng Wang, Fan Zhang, Yueze Wang, Zhen Li, Qiying Yu, et~al.
\newblock Emu3: Next-token prediction is all you need.
\newblock {\em arXiv preprint arXiv:2409.18869}, 2024.

\bibitem{wang2023paxion}
Zhenhailong Wang, Ansel Blume, Sha Li, Genglin Liu, Jaemin Cho, Zineng Tang, Mohit Bansal, and Heng Ji.
\newblock Paxion: Patching action knowledge in video-language foundation models.
\newblock {\em Advances in Neural Information Processing Systems}, 36:20729--20749, 2023.

\bibitem{yuksekgonul2022and}
Mert Yuksekgonul, Federico Bianchi, Pratyusha Kalluri, Dan Jurafsky, and James Zou.
\newblock When and why vision-language models behave like bags-of-words, and what to do about it?
\newblock {\em arXiv preprint arXiv:2210.01936}, 2022.

\bibitem{zhan2024anygpt}
Jun Zhan, Junqi Dai, Jiasheng Ye, Yunhua Zhou, Dong Zhang, Zhigeng Liu, Xin Zhang, Ruibin Yuan, Ge~Zhang, Linyang Li, et~al.
\newblock Anygpt: Unified multimodal llm with discrete sequence modeling.
\newblock {\em arXiv preprint arXiv:2402.12226}, 2024.

\end{thebibliography}
\appendix

\newpage

\section*{Appendix A: Detailed Model Formulation}
\label{append:formulation}


\subsection*{1. Modality-Specific Input Processing}

The model first processes the audio and text inputs in parallel to convert them into a shared embedding space.

\paragraph{Audio Processing}
A raw 10-second audio clip, $A_{\text{clip}}$, is fed into the pre-trained CAV-MAE audio encoder, $E_A$, which generates a sequence of 32 feature-rich embeddings, $H_A$:
$$
H_A = E_A(A_{\text{clip}}) \quad \text{where } H_A \in \mathbb{R}^{32 \times d_{\text{audio}}}
$$
These embeddings (where $d_{\text{audio}}=768$ for CAV-MAE) are then passed through a linear projection layer, $P$, to match the 4096-dimensional hidden space of the LLaMa backbone, resulting in the final audio sequence $Z_A$:
$$
Z_A = P(H_A) \quad \text{where } Z_A \in \mathbb{R}^{32 \times 4096}
$$

\paragraph{Text Processing}
A raw text prompt, $T$, is first processed by the LLaMa tokenizer, which converts the string into a sequence of $L$ integer token IDs, $T_{\text{ids}}$:
$$
T_{\text{ids}} = \text{Tokenizer}(T) = \{\text{id}_1, \text{id}_2, \ldots, \text{id}_L\}
$$
These token IDs are then passed through the LLaMa model's own word embedding layer, $\text{Embed}(\cdot)$, to produce the final text embedding sequence $E_T$:
$$
E_T = \text{Embed}(T) = \{t_1, t_2, \ldots, t_L\} \quad \text{where } E_T \in \mathbb{R}^{L \times 4096}
$$


\subsection*{2. Sequence Construction}

Once $Z_A$ and $E_T$ are computed, they are combined into a single input sequence $I$. The construction of this sequence differs between the baseline model and our proposed method.

\paragraph{Non-Interleaved Construction (Baseline)}
In the conventional non-interleaved setup, as used by the original LTU model, the entire sequence of audio embeddings $Z_A$ is prepended to the text embeddings $E_T$. This creates a single sequence $I_{\text{non-interleaved}}$:
$$
I_{\text{non-interleaved}} = [Z_A ; E_T] = \{z_1, \ldots, z_{32}, t_1, \ldots, t_L\}
$$
Here, the model first processes all audio information before processing any text.

\paragraph{Interleaved Construction}
For our interleaved setup, the final input sequence $I_{\text{interleaved}}$ is constructed by inserting the audio embeddings $Z_A$ into the text embeddings $E_T$ at the position indicated by a conceptual placeholder $\langle\text{audio}\rangle$. For a prompt $T = (T_{\text{pre}}, \langle\text{audio}\rangle, T_{\text{post}})$, the corresponding embedding sequences $E_{\text{pre}}$ and $E_{\text{post}}$ are combined with $Z_A$:
$$
I_{\text{interleaved}} = [E_{\text{pre}} ; Z_A ; E_{\text{post}}] = \{t_1, \ldots, t_i, z_1, \ldots, z_{32}, t_{i+1}, \ldots, t_L\}
$$
Let this full input sequence of total length $N$ tokens be denoted $X = \{x_1, x_2, \ldots, x_N\}$, where each $x_n$ is a 4096-dimensional vector from either $E_T$ or $Z_A$.
\subsection*{3. Training Objective}
\label{sec:appendix_loss}
The model is trained using a standard autoregressive, next-token prediction objective with a cross-entropy loss. Let $X = \{x_1, \ldots, x_N\}$ be the input sequence and $Y = \{y_1, \ldots, y_N\}$ be the target tokens (i.e., the input sequence shifted by one). The total loss $\mathcal{L}$ is:
$$
\mathcal{L} = -\sum_{n=1}^{N} \log P(y_n | x_1, \ldots, x_n)
$$
During fine-tuning, we only want the model to predict text tokens, not audio embeddings. We apply a binary mask $m_n$ to the loss, where $m_n = 1$ if $x_n$ is a text token and $m_n = 0$ if $x_n$ is an audio token. The final masked loss $\mathcal{L}_{\text{masked}}$ is:
$$
\mathcal{L}_{\text{masked}} = -\sum_{n=1}^{N} m_n \cdot \log P(y_n | x_1, \ldots, x_n)
$$
In practice, this masking is commonly implemented by setting the target labels $y_n$ corresponding to audio tokens to $-100$, which is the standard ignore-index for the cross-entropy loss.

\newpage
\section*{Appendix B: Prompt Generation for Fine-Tuning Dataset}
\label{append:promptGen}

\subsection*{1. Prompt Generation}

To create our fine-tuning datasets, we developed a pipeline to reformulate the original LTU audio question-answering data into our desired interleaved format. This process used GPT-4.1-mini to generate a set of diverse prompts based on two question types: closed-ended and open-ended questions. As described in LTU\cite{gong2023listen}, the closed-ended questions includes two types of classification tasks, label classification and acoustic feature classification, which have a fixed output format and objective answers. The open-ended questions are free-form and were designed to train the LTU model in advanced audio reasoning and comprehension abilities. 
These questions span multiple tasks, including audio reasoning 
and audio scene analysis.

The primary objective of this pipeline was to convert the original non-interleaved prompts into a new format where the \texttt{[AUDIO]} placeholder token was naturally integrated within the instruction's text. The generated prompts should be grammatically correct, semantically equivalent to the originals, and linguistically diverse, avoiding simple, repetitive templates.
In order to encourage GPT-4.1-mini to vary its sentence structure and phrasing, a randomly selected instruction from a predefined list was appended to each system prompt (see Table \ref{tab:variety_prompts}). 

Moreover, a temperature sampling strategy with values ranging from 0.7 to 1.1 was employed during this generation process to encourage further variability in wording. Crucially, while the textual prompt was reformulated, the source audio clip and the corresponding ground-truth answer label remained identical to the original dataset, ensuring that the primary experimental variable was the prompt format itself. 

\begin{table*}[h!]
\caption{List of instructions randomly selected from and appended to the system prompt during generation to increase linguistic diversity.}
\label{tab:variety_prompts}
\begin{tabularx}{\textwidth}{@{}X@{}}
\toprule

\begin{itemize}
    \item \textit{Use creative and varied language.}
    \item \textit{Employ different sentence structures and word choices.}
    \item \textit{Be innovative in your phrasing while maintaining clarity.}
    \item \textit{Use diverse vocabulary and avoid repetitive patterns.}
    \item \textit{Create unique formulations while keeping the core meaning.}
    \item \textit{Vary your word choice and sentence construction.}
    \item \textit{Express the same concept using different linguistic approaches.}
    \item \textit{Be original in your expression while preserving the instruction's purpose.}
\end{itemize}
\\
\bottomrule
\end{tabularx}
\end{table*}

The exact prompt templates used to guide the generation for each task are detailed in Table \ref{tab:label_prompt}, Table \ref{tab:description_prompt}, and Table \ref{tab:open_ended_prompt}. Each prompt template consists of two primary parts: a system prompt that defines the LLM's role and general task rules, and a user prompt that provides the specific task. As shown in the tables, these templates contain placeholders (\texttt{\{non-interleaved prompt\}} and \texttt{\{instruction\}}). 
The \texttt{\{non-interleaved prompt\}} is filled with a non-interleaved prompt used in the original LTU, while \texttt{\{instruction\}} is filled with a randomly selected instruction from Table \ref{tab:variety_prompts} to ensure linguistic diversity.
While all templates share common rules (e.g., the mandatory single \texttt{[AUDIO]} placeholder or the instruction to avoid referring to media files directly) in the system prompt, each includes unique task-specific constraints. For example, the prompt for acoustic feature classification (Table \ref{tab:description_prompt}) requires the model to ask for both a label and its acoustic properties, while the open-ended QA prompt (Table \ref{tab:open_ended_prompt}) focuses on rephrasing the original question. We employed a few-shot prompting technique to present a small number of good example responses to the LLM, ensuring that the generated interleaved prompts are of high quality, task-appropriate, and aligned with our research goals.


\newpage
\begin{table}[H]
\caption{Input prompt template for reformulating non-interleaved prompts into interleaved prompts for \textbf{closed-ended label classification} task. A random instruction from Table \ref{tab:variety_prompts} was appended to the System prompt to diversify the output.}
\label{tab:label_prompt}
\centering
\begin{tabularx}{\linewidth}{@{}lX@{}}
\toprule
\textbf{Role} & \textbf{Content} \\
\midrule
\textbf{System} &
You are an expert AI assistant specializing in revising prompts for multimodal language models.
Your task is to rewrite a given prompt into a new, interleaved format.
\newline\newline
\textbf{Your Rules:}
\begin{enumerate}
    \item You must take the user's 'Old Prompt' and rephrase it into an abstract, interleaved instruction.
    \item The new prompt must contain the exact placeholder \texttt{[AUDIO]} one and only one time.
    \item The new prompt must avoid words that explicitly refer to a media file, such as "clip," "recording," or "audio file."
    \item The new prompt must be completely general and scenario-agnostic.
    \item Your final output must be a single JSON object with one key: "revised\_prompt". Do not include any other text.
\end{enumerate}
\vspace{0.5em}
IMPORTANT: \texttt{\{instruction\}}  Make each instruction distinct and avoid formulaic responses. Use different words and sentence structures even when the meaning is similar.
\\ \midrule
\addlinespace[10pt]
\textbf{User} &
I need to revise the following prompt for a simple audio classification task. The goal is to ask for a list of labels.
\newline\newline
\textbf{Old Prompt:} "\texttt{\{non-interleaved prompt\}}"
\newline\newline
Please revise it into a new, single-string interleaved prompt.
\newline\newline
\textbf{Good Revision Examples:}
\begin{itemize}
    \item "After you hear \texttt{[AUDIO]}, what are the appropriate classification labels?"
    \item "Listen to this: \texttt{[AUDIO]}. Now, list the corresponding tags."
    \item "Consider \texttt{[AUDIO]}. What are its corresponding labels?"
\end{itemize}
\vspace{0.5em}
Provide your output as a single JSON object with the key "revised\_prompt".
\\
\bottomrule
\end{tabularx}
\end{table}

\begin{table}[H]
\caption{Input prompt template for reformulating non-interleaved prompts into interleaved prompts for the \textbf{acoustic feature classification} fine-tuning task. A random instruction from Table \ref{tab:variety_prompts} was appended to the System prompt to diversify the ouptut.}
\label{tab:description_prompt}
\centering
\begin{tabularx}{\linewidth}{@{}lX@{}}
\toprule
\textbf{Role} & \textbf{Content} \\
\midrule
\textbf{System} &
You are an expert AI assistant specializing in revising prompts for multimodal language models.
Your task is to rewrite a given prompt into a new, interleaved format for a complex audio classification task that requires acoustic descriptions.
\newline\newline
\textbf{Your Rules:}
\begin{enumerate}
    \item You must take the user's 'Old Prompt' and rephrase it into an abstract, interleaved instruction.
    \item The new prompt must contain the exact placeholder \texttt{[AUDIO]} one and only one time.
    \item The prompt must explicitly ask for both a label AND a description of its acoustic features.
    \item The new prompt must avoid words that explicitly refer to a media file, such as "clip," "recording," or "audio file."
    \item The new prompt must be completely general and scenario-agnostic.
    \item Your final output must be a single JSON object with one key: revised\_prompt, e.g. \texttt{\{'revised\_prompt': '...'\}}. Do not include any other text.
\end{enumerate}
\vspace{0.5em}
IMPORTANT: \texttt{\{instruction\}}  Make each instruction distinct and avoid formulaic responses. Use different words and sentence structures even when the meaning is similar.
\\ \midrule
\addlinespace[10pt]
\textbf{User} &
I need to revise the following prompt for a complex audio classification task. The goal is to ask for a list of labels, each with a description of its acoustic properties.
\newline\newline
\textbf{Old Prompt:} "\texttt{\{non-interleaved prompt\}}"
\newline\newline
Please revise it into a new, single-string interleaved prompt.
\newline\newline
\textbf{Good Revision Examples:}
\begin{itemize}
    \item "Listen to this: \texttt{[AUDIO]}. For each component you identify, list its label and describe its acoustic features."
    \item "Regarding \texttt{[AUDIO]}, what labels are suitable, and what are their key sound properties?"
    \item "Analyze what you hear in \texttt{[AUDIO]}. Return a list of labels paired with their distinguishing acoustic qualities."
\end{itemize}
\vspace{0.5em}
Provide your output as a single JSON object with the key revised\_prompt, e.g. \texttt{\{'revised\_prompt': '...'\}}.
\\
\bottomrule
\end{tabularx}
\end{table}

\begin{table}[H]
\caption{Input prompt template for reformulating non-interleaved prompts into interleaved prompts for the \textbf{open-ended QA} fine-tuning task. A random variety prompt from Table \ref{tab:variety_prompts} was appended to the System prompt to diversify the output.}
\label{tab:open_ended_prompt}
\centering
\begin{tabularx}{\linewidth}{@{}lX@{}}
\toprule
\textbf{Role} & \textbf{Content} \\
\midrule
\textbf{System} &
You are an expert AI assistant specializing in revising prompts for multimodal language models.
Your task is to rewrite a given open-ended question into a new, interleaved format.
\newline\newline
\textbf{Your Rules:}
\begin{enumerate}
    \item You must take the user's 'Old Prompt' and rephrase it by naturally integrating the \texttt{[AUDIO]} placeholder into the question.
    \item The new prompt must preserve the full intent and meaning of the original question.
    \item The new prompt must contain the exact placeholder \texttt{[AUDIO]} one and only one time.
    \item The new prompt must avoid words that explicitly refer to a media file, such as "clip," "recording," or "audio file."
    \item The resulting prompt should be a single, grammatically correct, and natural-sounding question.
    \item Your final output must be a single JSON object with one key: "revised\_prompt". Do not include any other text.
\end{enumerate}
\vspace{0.5em}
IMPORTANT: \texttt{\{instruction\}}  Make each instruction distinct and avoid formulaic responses. Use different words and sentence structures even when the meaning is similar.
\\ \midrule
\addlinespace[10pt]
\textbf{User} &
I need to revise the following prompt for an open-ended audio question-answering task. The goal is to rephrase the question to include an audio placeholder.
\newline\newline
\textbf{Old Prompt:} "\texttt{\{non-interleaved prompt\}}"
\newline\newline
Please revise it into a new, single-string interleaved prompt.
\newline\newline
\textbf{Good Revision Examples:}
\begin{itemize}
    \item Old: "What other sound events, if any, can be heard in the audio clip?" -> New: "What other sound events, if any, can be heard in \texttt{[AUDIO]}?"
    \item Old: "Based on the acoustic features, can you tell the type of vacuum cleaner?" -> New: "Based on \texttt{[AUDIO]}, can you tell the type of vacuum cleaner?"
    \item Old: "Describe the environment where this sound was likely recorded." -> New: "Based on what you hear in \texttt{[AUDIO]}, describe the environment where it was likely recorded."
\end{itemize}
\vspace{0.5em}
Provide your output as a single JSON object with the key "revised\_prompt".
\\
\bottomrule
\end{tabularx}
\end{table}

\subsection*{2. Sample Output}


Below are representative examples from the resulting fine-tuning dataset, showing the original (non-interleaved) prompt, the new interleaved reformulation, and the ground-truth answer. These examples are based on the original LTU audio question-answering data, which consists of two main types: closed-ended and open-ended questions. 
The closed-ended questions include classification tasks like simple label classification (Table \ref{tab:prompt_examples1}) and more complex acoustic feature classification (Table \ref{tab:prompt_examples2}). The open-ended questions (Table \ref{tab:prompt_examples3}) are free-form and designed to train advanced audio reasoning and comprehension abilities.

\begin{table*}[h!]
\caption{Sample Example: Classification Task (Label)}
\label{tab:prompt_examples1}

\begin{tabularx}{\textwidth}{@{}X@{}}
\toprule
    \paragraph{Original Prompt:}
    Analyze audio events in clip given.
    
    \paragraph{Interleaved Prompt:}
    Reflect on the contents of \texttt{[AUDIO]} and enumerate the relevant categories it represents.
    
    \paragraph{Ground-Truth Answer:}
    Labels: Squeal
\\
\bottomrule
\end{tabularx}
\end{table*}


\begin{table*}[h!]
\caption{Sample Example: Classification Task (Acoustic Features)}
\label{tab:prompt_examples2}
\begin{tabularx}{\textwidth}{@{}X@{}}
\toprule
    \paragraph{Original Prompt:}
    Identify the noise in the audio clip? Analyze acoustic features first.
    
    \paragraph{Interleaved Prompt:}
    Consider \texttt{[AUDIO]} and enumerate all discernible sound categories, specifying for each both an appropriate label and a detailed account of its auditory characteristics.
    
    \paragraph{Ground-Truth Answer:}
    Labels with acoustic features: Punctuated and deep -> Car; High-pitched, screeching and piercing -> Tire squeal; Sharp and abrasive -> Skidding; Loud, deep and metallic -> Vehicle
\\
\bottomrule
\end{tabularx}
\end{table*}

\newpage
\begin{table}[H]
\caption{Sample Example: Open-ended Task (Description)}
\label{tab:prompt_examples3}
\begin{tabularx}{\textwidth}{@{}X@{}}
\toprule
    \paragraph{Original Prompt:}
    What is the volume of the electric shaver sound compared to the music and singing?
    
    \paragraph{Interleaved Prompt:}
    How does the volume of the electric shaver sound in \texttt{[AUDIO]} compare to the music and singing?
    
    \paragraph{Ground-Truth Answer:}
    The electric shaver sound is described as high-pitched and faint compared to the clear and vibrant music and rich and resonant singing.
\\
\bottomrule
\end{tabularx}
\end{table}

\section*{Appendix C: Disclosure on the Use of LLMs}
\label{sec:appendix_llm_use}

First, an LLM was an integral component of our fine-tuning data generation pipeline. We used GPT-4.1-mini to systematically reformulate prompts from the original LTU dataset into the interleaved format used for our experiments. The specific prompts used to guide this generation process are detailed in Appendix B.

Second, we used OpenAI's GPT-5 to help refine code and improve grammar as well as clarity of the manuscript. The scientific concepts, experimental design, results, and conclusions presented in this paper are the original work of the human authors, who take full responsibility for the content of the paper.


\end{document}